\documentclass{Interspeech}

\interspeechcameraready

\title{Leveraging Mamba with Full-Face Vision for Audio-Visual Speech Enhancement}

\author[affiliation={1, 2}]{Rong}{Chao}
\author[affiliation={1, 2}]{Wenze}{Ren}
\author[affiliation={1, 2}]{You-Jin}{Li}
\author[affiliation={1, 2}]{Kuo-Hsuan}{Hung}
\author[affiliation={3}]{Sung-Feng}{Huang}
\author[affiliation={3}]{Szu-Wei}{Fu}
\author[affiliation={2}]{Wen-Huang}{Cheng}
\author[affiliation={1}]{Yu}{Tsao}

\affiliation{CITI}{Academia Sinica}{Taiwan}
\affiliation{CSIE}{National Taiwan University}{Taiwan}
\affiliation{}{NVIDIA}{}
\email{d13922037@ntu.edu.tw, yu.tsao@citi.sinica.edu.tw}

\keywords{speech enhancement, audio-visual speech enhancement, AVSEC-4 Challenge, state-space models, Mamba}

\usepackage{comment}
\usepackage{amsmath,graphicx}
\usepackage{booktabs}
\usepackage{graphicx}
\usepackage{xcolor}
\usepackage{amsmath, amssymb, graphicx}
\usepackage{algorithm}
\usepackage{algpseudocode}
\usepackage{url}
\usepackage{ragged2e}
\usepackage{balance}
\usepackage{multirow}
\usepackage{amssymb}
\usepackage{subfig}
\usepackage{makecell} 

\usepackage{booktabs}

\begin{document}
%
\setlength{\lightrulewidth}{0.02em}
\setlength{\heavyrulewidth}{0.08em}
%
\setlength{\textfloatsep}{5.0pt plus 3.0pt minus 3.0pt}
\setlength{\floatsep}{5.0pt plus 3.0pt minus 3.0pt}
\setlength{\intextsep}{3.0pt plus 3.0pt minus 3.0pt}
\setlength{\dbltextfloatsep}{6.0pt plus 3.0pt minus 2.0pt}

\maketitle


\begin{abstract}
Recent Mamba-based models have shown promise in speech enhancement by efficiently modeling long-range temporal dependencies. However, models like Speech Enhancement Mamba (SEMamba) remain limited to single-speaker scenarios and struggle in complex multi-speaker environments such as the cocktail party problem. To overcome this, we introduce AVSEMamba, an audio-visual speech enhancement model that integrates full-face visual cues with a Mamba-based temporal backbone. By leveraging spatiotemporal visual information, AVSEMamba enables more accurate extraction of target speech in challenging conditions. Evaluated on the AVSEC-4 Challenge development and blind test sets, AVSEMamba outperforms other monaural baselines in speech intelligibility (STOI), perceptual quality (PESQ), and non-intrusive quality (UTMOS), and achieves \textbf{1st place} on the monaural leaderboard.
\end{abstract}




\section{Introduction}

The cocktail party problem, which involves isolating a target speaker in noisy, multi-speaker environments, remains a fundamental challenge in speech enhancement. Human listeners excel in such conditions by naturally integrating auditory and visual cues such as lip motion and facial expressions. Audio-Visual Speech Enhancement (AVSE) systems aim to replicate this multisensory processing to improve intelligibility under adverse acoustic conditions.

AVSE methods have evolved notably in recent years. Early works focused on lip motion to guide enhancement \cite{afouras2018deep, ephrat2018looking}, while later studies addressed robustness to occlusions and facial variance \cite{afouras2019my}. Subsequent efforts expanded visual input from the mouth region to the full face \cite{chung2020facefilter}, and incorporated SSL-based cross-modal consistency constraints for better alignment~\cite{chern2023audio}. More recently, DCUC-Net~\cite{ahmed2023deep} leveraged a Conformer-based architecture and scene-level visual cues beyond facial regions to enhance separation quality.

While recent AVSE systems have shown strong performance, most architectures rely on attention-based mechanisms that may struggle with scalability in long sequences. Given the efficiency and favorable memory scaling of state space models, we investigate AVSEMamba. This hybrid AVSE system integrates full-face video and audio features through Mamba-based temporal-frequency modeling. 
Our approach is designed to handle visually guided speech enhancement under multi-speaker conditions, and we validate its effectiveness on the AVSEC-4 Challenge benchmark with consistent improvements in intelligibility, perceptual quality, and ASR accuracy.

\section{Related Work}
In terms of temporal modeling, early AVSE systems employed CNNs and RNNs, which were later replaced by Transformer-based architectures for enhanced sequence modeling. However, Transformers scale quadratically with input length, limiting their suitability for long-context or real-time applications.



State space models (SSMs) offer an efficient alternative with linear complexity. Among them, Mamba introduces input-conditioned selective scanning, enabling scalable long-range modeling. SAV-Net \cite{qian2025sav} introduced Mamba into audio-visual speech enhancement (AVSE) using scene-level visual cues. Separately, SEMamba and its extension USEMamba \cite{chao2024investigation, chao2025universal} applied Mamba to monaural enhancement with low compute cost. Building on this line, AVSEMamba incorporates full-face visual input into a Mamba-based AVSE framework, enabling robust target speech extraction in multi-speaker scenarios.


\begin{figure*}[t]
    \centering
    \includegraphics[width=0.85\textwidth]{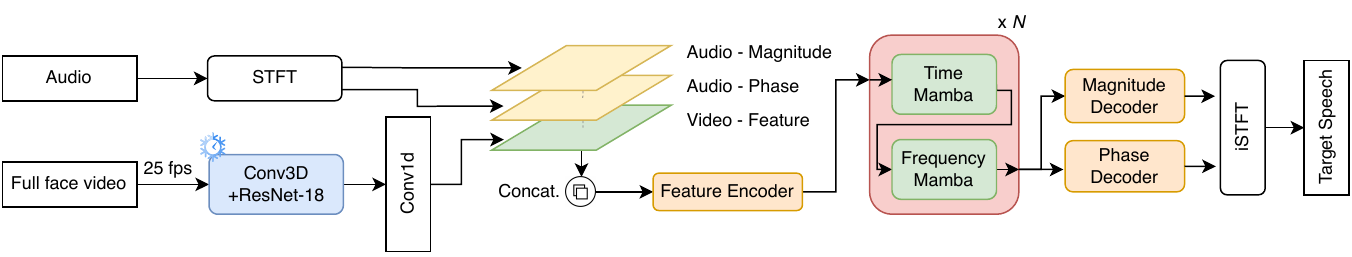}
    \caption{System architecture of the proposed AVSEMamba model.}
    \label{fig:avsemamba}
\end{figure*}

\section{Audio-Visual Speech Enhancement with Mamba}

We propose a hybrid AVSEMamba architecture that extends the SEMamba \cite{chao2024investigation} framework by integrating full-face visual information for improved performance in multi-speaker scenarios. The model leverages Mamba-based state space modeling to capture long-range dependencies across both temporal and frequency dimensions.

The system consists of two synchronized input streams:
\begin{itemize}
	\item Audio Stream: Raw waveforms are transformed via Short-Time Fourier Transform (STFT), producing magnitude and phase components, each processed by a convolutional frontend.
	\item Visual Stream: Full-face video (25 fps) is passed through a pretrained and frozen 3D ResNet-18 to extract spatiotemporal embeddings. These are temporally aligned with the audio via a Conv1D layer.

\end{itemize}

The audio magnitude, phase, and visual features are concatenated and passed through a shared Feature Encoder, which projects them into a common representation space. This fused representation is then fed into a stack of Mamba-based temporal-frequency modeling blocks, which operate jointly across time and frequency axes. Finally, the network predicts enhanced magnitude and phase components through separate decoders, and the target speech waveform is reconstructed using inverse STFT, as illustrated in Fig.~\ref{fig:avsemamba}.

By leveraging Mamba’s structured sequence modeling and full-face visual context, our system captures both spatial-temporal and cross-modal dependencies, resulting in robust enhancement performance under occlusion, reverberation, and acoustic mismatch.

\section{Experiments}

We evaluate our proposed AVSEMamba model on the AVSEC-4 Challenge dataset using \textbf{monophonic audio input and prediction only}. The dataset comprises synchronized full-face video and audio data under various acoustic and visual interference conditions. All evaluations are conducted using mono audio, in contrast to the challenge’s default binaural setup.

\vspace{-0.7em}
\subsection{Dataset}

The AVSEC-4 training set comprises 34,524 mono scenes with 605 target speakers and interferers, sampled from 405 competing speakers, and 7,346 noise clips spanning 15 categories, including speech, non-speech (e.g., domestic and environmental), and music. Target speech is derived from the LRS3 dataset.

The development set consists of 3,306 mono scenes (approximately 8 hours) with 85 target speakers and similar interferer sources. Audio is sampled at 16\,kHz with 16-bit depth. 


\vspace{-0.7em}
\subsection{Training Strategy}

Models are trained for 450 epochs using the AdamW optimizer with a batch size of 2. The loss function is a weighted combination of waveform $L_1$ loss, STFT magnitude loss, complex spectral loss, phase loss, and consistency loss.

\subsection{Results on Development Set}

As shown in Table~\ref{tab:dev}, our AVSEMamba achieves notably improvements over the noisy input and the Challenge-provided monaural baseline. Specifically, our model improves MBSTOI by \textbf{93.1}\% (from 0.4161 to 0.8037) and PESQ by \textbf{128.5}\% (from 1.30 to 2.97). Notably, the UTMOS score of our model is also very close to that of the clean target (2.32 vs. 2.36).

\vspace{1em}
\begin{table}[h]
\centering
\caption{Performance on the AVSEC-4 development set (mono audio). Note: mono audio was duplicated for MBSTOI.}
\label{tab:dev}
\begin{tabular}{lccc}
\toprule
\textbf{Method} & \textbf{MBSTOI} & \textbf{PESQ} & \textbf{UTMOS} \\
\midrule
Clean target    & - & - & 2.36 \\
\cmidrule(lr){1-4}
Noisy Input     & 0.4161 & 1.30 & 1.32 \\
Baseline (Mono)  & - & 1.21 & 1.26 \\
Ours (AVSEMamba) & \textbf{0.8037} & \textbf{2.97} & \textbf{2.32} \\
\bottomrule
\end{tabular}
\end{table}

\begin{table}[h]
\centering
\caption{Results on the AVSEC-4 Blind Test Set (Mono Audio).}
\label{tab:blind}
\begin{tabular}{lccc}
\toprule
\textbf{Method} & \textbf{PESQ} & \textbf{STOI} & \textbf{UTMOS} \\
\midrule
Noisy Input      & 1.31 & 0.55 & 1.37 \\
Baseline (Mono)  & 1.25 & 0.33 & 1.26 \\
Ours (AVSEMamba) & \textbf{2.31} & \textbf{0.77} & \textbf{2.24} \\
\bottomrule
\end{tabular}
\end{table}

\subsection{Results on Blind Test Set}


Table~\ref{tab:blind} summarizes the performance on the AVSEC-4 blind test set. PESQ and STOI scores are reported by the official leaderboard (evaluated on a subset), while UTMOS was computed on the full test set.
Compared to the noisy input, AVSEMamba achieves a \textbf{76.3\%} improvement in PESQ (1.31 → 2.31), a \textbf{40.0\%} gain in STOI (0.55 → 0.77), and a \textbf{63.5\%} increase in UTMOS (1.37 → 2.24).

\vspace{-0.7em}
\section{Conclusion}




We presented AVSEMamba, a hybrid audio-visual speech enhancement architecture that integrates full-face visual cues with Mamba-based temporal-frequency modeling. By leveraging pre-trained spatiotemporal visual features and efficient state-space modeling, AVSEMamba effectively addresses visually guided enhancement in multi-speaker conditions.

Evaluated on the AVSEC-4 Challenge, AVSEMamba achieved strong performance across all metrics and ranked first in PESQ on the official leaderboard, demonstrating the potential of state-space models for audio-visual speech enhancement.

\vspace{-0.7em}

\section{Acknowledgment}
This work was supported in part by the NVIDIA Academic Grant Program. The authors gratefully acknowledge NVIDIA Corporation for providing GPU resources used in this research.

\vspace{-0.7em}

\bibliographystyle{IEEEtran}
\bibliography{mybib}

\end{document}